# Physics based calculation of the fine structure constant

J. P. Lestone
Applied Physics Division, Los Alamos National Laboratory,
Los Alamos, New Mexico 87545, USA
(February 5$^{th}$ 2009)

**Abstract**

We assume that the coupling between particles and photons is defined by a surface area and a temperature, and that the square of the temperature is the inverse of the surface area ($\hbar=c=1$). By making assumptions regarding stimulated emission and effects associated with the finite length of a string that forms the particle surface, the fine structure constant is calculated to be ~1/137.04. The corresponding calculated fundamental unit of charge is $1.6021 \times 10^{-19}$ C.

**Introduction**

Richard Feynman called the value of the fine structure constant [1] "one of the greatest damn mysteries of physics." Decades later, there is still no accepted theory to explain the value of the fine structure constant $\alpha \sim 1/137$. In the present paper we assume that particles have an effective emission surface area and temperature. We assume that the square of the temperature is the inverse of the surface area ($\hbar=c=1$). After making some assumptions about the nature of stimulated emission from particles, and effects associated with the finite length of a string that forms the particle surface, we calculate the repulsive force associated with the exchange of photons between two electrons. This photon exchange force gives a calculated fine structure constant close to 1/137.

**Model**

We assume that the photon emission and absorption area $A$ of an electron is controlled by a length scale $f$. We further assume that the electron has a corresponding effective mean temperature $T$ and that the relationship between $T$ and $f$ is the same as the relationship between the Planck length $L_P$ and the Planck temperature $T_P$, and thus we write

$$T^2 = \frac{L_P^2 T_P^2}{f^2} = \frac{1}{A} \frac{\hbar G}{c^3} \frac{\hbar c^5}{G} = \frac{(\hbar c)^2}{A}. \qquad (1)$$

This temperature is $4\pi$ times larger than the Hawking temperature of a Schwarzschild black-hole [2] with a radius $f$. Even though we assume an effective temperature, a single electron is incapable of radiating to infinity because this would violate conservation of energy and momentum. However, for a system consisting of two electrons we assume that photons can be exchanged, and that the strength of this exchange is controlled by the emission and absorption of black-body radiation. If we ignore that photon absorption by an object having a temperature should have a corresponding stimulated emission process, then the force generated by the exchange of photons between two electrons is given by

$$F = \frac{2A\sigma T^4}{c} \frac{A/4}{4\pi d^2} = \frac{\pi}{480} \frac{\hbar c}{d^2} = \frac{q^2}{4\pi\varepsilon_o d^2}, \qquad (2)$$

where $d$ is the distance between the particles. Eq. (2) gives a value for the inverse fine structure constant of $\alpha^{-1} = (4\pi\varepsilon_0\hbar c)/q^2 = 480/\pi \sim 152.8$. This is ~1.1 times the value of the real inverse fine structure constant $\alpha^{-1} = 137.035999070(98)$ [3]. The corresponding calculated effective charge is

$$q = \sqrt{\frac{4\pi\varepsilon_0 \hbar c}{152.8}} \sim 1.52 \times 10^{-19} \text{ C}. \tag{3}$$

If we proceeded no further, this result could be dismissed as a coincidence.

In analogy with other absorption processes, we assume that the absorption cross section $A/4$ should be associated with a corresponding stimulated emission cross section $A/4 \, e^{-\varepsilon}$ where $\varepsilon$ is the energy of the incident photon relative to the temperature of the system. We assume that when a photon is absorbed by an electron, there is a probability $e^{-\varepsilon}$ that a stimulated emission occurs. As discussed earlier, we assume that there is no emission to infinity and thus assume that the stimulated photon retraces the incident photon's path and is re-absorbed by the other electron. This re-absorption of the stimulated photon has a probability $e^{-\varepsilon}$ of generating another stimulated emission, and so on. Therefore, for each spontaneously emitted photon exchanged between two electrons, the total number of exchanged photons (both spontaneous and stimulated) is

$$N(\varepsilon) = \sum_{k=0}^{\infty} \exp(-k\varepsilon) = \frac{\exp(\varepsilon)}{\exp(\varepsilon) - 1}. \tag{4}$$

This modifies the force presented in Eq. (2) to

$$F = \frac{\pi}{480} \frac{\hbar c}{d^2} \int_0^\infty \frac{15}{\pi^4} \frac{\varepsilon^3 \exp(\varepsilon)}{(\exp(\varepsilon)-1)^2} d\varepsilon. \tag{5}$$

This gives an inverse fine structure constant of the exchange of photons between two electrons

$$\alpha^{-1} = \frac{480}{\pi} \div \int_0^\infty \frac{15}{\pi^4} \frac{\varepsilon^3 \exp(\varepsilon)}{(\exp(\varepsilon)-1)^2} d\varepsilon = \frac{480}{\pi} \frac{\pi^4}{15 \cdot 6 \cdot \zeta(3)} = \frac{16\pi^3}{3\sum_{k=1}^{\infty} k^{-3}} = \frac{16\pi^3}{3\prod_{p \text{ prime}} \frac{1}{1-p^{-3}}} \sim 137.5699. \tag{6}$$

$\zeta(n)$ is the Riemann zeta-function. The result given in Eq. (6) is in disagreement with the corresponding measured value [3] by ~4 parts in a 1000. The corresponding calculated fundamental unit of charge is $q \sim 1.60 \times 10^{-19}$ C. These results are close enough to the corresponding experimental values that it is more difficult to dismiss them as being obtained via pure coincidence. Given that the result in Eq. (6) is related to the temperature to the 4th power, good agreement with the measured value for $\alpha$ could be obtained if the assumed temperature is increased by ~1 part in a 1000.

To add further complexity, we assume that an electron consists of a loop of string with length $\ell_s$ moving about on the 2-dimensional surface of a nearly spherical membrane with radius $f$. We further speculate that the string's length is $n$ times the sphere's circumference, and long enough that, in a short time interval, it can effectively cover most of the sphere's surface, and we thus assume $\ell_s = n2\pi f$. We suggest that $n=3$ is the smallest integer that is likely to satisfy the above conditions. We assume that the finite length of the string generates an uncertainty in the effective temperature of the particle and that this temperature uncertainty is equal to the energy uncertainty related to the time it takes a signal to travel the length of the string $\Delta E = \hbar c/(2\ell_s)$. We therefore write

$$\sigma_T = \frac{\Delta T}{T} = \frac{\hbar c}{2 \times n2\pi f} \frac{f}{\hbar c} = \frac{1}{n4\pi}. \tag{7}$$

The inverse fine structure constant associated with the exchange of photons between two electrons can then be expressed as

$$\alpha^{-1} = \frac{16\pi^3}{3 \cdot \zeta(3)} \frac{1}{g(\sigma_T)}, \tag{8}$$

where $g(\sigma_T)$ is the string correction factor,

$$g(\sigma_T) = \lim_{j \to \infty} \int\int\int...\int_{i=1,2,3,...}^{j} \frac{e^{(-\sum_{i=1}^{j} \Delta_i^2/(2\sigma_T^2))}}{(\sigma_T\sqrt{2\pi})^j} \int_{\varepsilon=0}^{\infty} \frac{\varepsilon^3}{e^{\varepsilon/(1+\Delta_1)}-1}(1+e^{-\varepsilon/(1+\Delta_2)}(1+e^{-\varepsilon/(1+\Delta_3)}\times$$
$$(1+...\,e^{-\varepsilon/(1+\Delta_j)}\frac{e^\varepsilon}{e^\varepsilon-1}))) \frac{d\varepsilon d\Delta_{1..j}}{6\cdot\zeta(3)}. \tag{9}$$

This is a difficult integral to solve exactly. Ignoring the possibility of an analytical solution, obtaining $g(\sigma_T)$ to the accuracy of 9 or more significant digits using Eq. (9) would require numerically solving a several-hundred dimensional integral. This is not possible. However, given that $\sigma_T \ll 1$, Eq. (9) can be approximated by a product of an infinite number of one dimensional integrals. By numerically solving Eq. (9) with $\sigma_T \sim 1/(12\pi)$ and $j$=1 to 5, one can deduce that

$$g(\sigma_T) \sim 1.00000014 \times \prod_{i=1}^{\infty} \int_{\Delta=-\infty}^{\infty} \frac{e^{-\Delta_i^2/(2\sigma_T^2)}}{\sigma_T\sqrt{2\pi}} \int_{\varepsilon=0}^{\infty} \frac{\varepsilon^3}{e^{\varepsilon/(1+\Delta_1\delta_{1i})}-1}(1+e^{-\varepsilon/(1+\Delta_2\delta_{2i})}\times$$
$$(1+e^{-\varepsilon/(1+\Delta_3\delta_{3i})}(1+e^{-\varepsilon/(1+\Delta_4\delta_{4i})}(1+...))))\frac{d\varepsilon d\Delta}{6\cdot\zeta(3)}. \tag{10}$$

By solving Eq. (10) with $i$ to several hundred and extrapolating to $i=\infty$, we have determined that $n$=2.980192 gives a calculated fine structure constant that is consistent with the measured value. We suggest that for reasons presently unknown, the correct value for $n$ is exactly 3 and that other higher order effects may lead to agreement with experiment. Perhaps the value of $n$=3 is related to the 3 color charges, the 1/3 fractional charges of quarks, the 3 generations of elementary particles, or 3 dimensional space. Assuming $n$=3 gives $g(\sigma_T=1/(12\pi))$=1.00384455 and $\alpha^{-1}$=137.042999. The corresponding calculated fundamental unit of charge is $q$=1.602136×10$^{-19}$ C. If we assume that the string length is either 2 or 4 times the circumference of the sphere then the corresponding calculated inverse fine structure constants are 136.390 and 137.273, respectively. Assuming $n$=3, good agreement with the measured value for $\alpha^{-1}$ could be obtained if the assumed effective particle temperature is increased by ~1 part in $10^5$ or by increasing $\sigma_T$ by ~1 part in a 100.

Assuming that there is a higher order correction to the temperature uncertainty, we express $\sigma_T$ as $s/(12\pi)$. One can easily confirm that, for small changes in $\sigma_T$, the corresponding corrected string correction factor can be expressed as

$$g(\sigma_T) = 1 + 0.00384455 \times s^2. \tag{11}$$

Without additional physical guidance, a large number of guesses for the functional form of $s$ can be found that give values of $\alpha^{-1}$ that are much closer to experiment than the value of 137.043 presented above. For example,

$$s = (\sum_{n=1}^{\infty} \frac{n^3}{(16\times 12\pi)^{n-1}})^{1/2} \text{ gives } \alpha^{-1}\text{=137.035999}. \tag{12}$$

This highlights the near futility of speculating further without additional physical guidance from a detailed string based model.

**Discussion**

We have used relatively simple physics concepts in an attempt to understand the value of $\alpha^{-1}$. Perhaps the most novel concept described here is the assumption that electrons have both a surface area and an effective temperature that are coupled. The corresponding calculated $\alpha^{-1}$ is within ~10% of the known value. This result could be easily dismissed as a coincidence. However, by assuming an effective temperature, it is logical to then assume that particles have internal degrees of freedom. This, in turn, leads to the possibility of corrections associated with stimulated emission and string length. After applying estimates of these corrections, we obtained $\alpha^{-1}$~137.04.

When trying to explain or calculate the value of the fine structure constant, it is difficult to separate reasonable assumptions from assumptions that are chosen to give a result close to the known desired value. Serious concerns with the present work include the assumption that the photon capture cross section is independent of photon energy; that the stimulated emission is assumed to perfectly retrace the path of the inducing incoming photon; and that, for electrons to have a size smaller than ~$10^{-18}$ m, the effective temperature would be very much larger than the rest-mass energy of the electron. Despite these concerns, the fine structure constant calculated here suggests that the forces between fundamental particles are due to the exchange of bosons between particles having both a surface area and an effective temperature; and that the internal structure of electrons is string-like with an internal length scale close to 3 times the particle's circumference.

**Acknowledgments**

This work was supported in part by the U. S. Department of Energy.